\documentclass[runningheads]{llncs}
\usepackage[T1]{fontenc}
\usepackage{graphicx}
\begin{document}
\title{Adenocarcinoma Segmentation Using Pre-trained Swin-UNet with Parallel Cross-Attention for Multi-Domain Imaging}
\author{Abdul Qayyum\inst{1,2} \and Moona Mazher\inst{3} Imran Razzak\inst{4} 
 \and Steven A Niederer\inst{1,2} }
\authorrunning{A. Qayyum et al.}
\institute{National Heart \& Lung Institute, Faculty of Medicine, Imperial College London, United Kingdom \and
Turing Research and Innovation Cluster: Digital Twins, The Alan Turing Institute, London, United Kingdom \and
Department of Computer Science, University College London, United Kingdom\\
 \and
School of Computer Science and Engineering, University of New South Wales, Sydney, Australia \\
\email{\{a.qayyum, s.niederer\}@imperial.ac.uk} \\
\email{m.mazher@ucl.ac.uk, imran.razzak@unsw.edu.au}
}
\maketitle              % typeset the header of the contribution
\begin{abstract}
% Digital pathology has advanced significantly through AI and machine learning, improving tumor diagnosis and tissue segmentation. However, domain-shift—caused by variability in anatomical structures, tissue preparation, and imaging processes—challenges the robustness of segmentation models. 

% To address this, we propose a method using a pre-trained encoder on the ImageNet dataset with a Swin-UNet architecture enhanced by a parallel cross-attention module. Evaluated for Cross-Organ and Cross-Scanner Adenocarcinoma Segmentation, our model achieved final ranking scores of 0.7469 for Task 1 and 0.7597 for Task 2. By leveraging pre-training and advanced data augmentation, our approach effectively navigates diverse imaging conditions and improves segmentation accuracy across varying domains. These results demonstrate that our solution is robust and enhances medical image analysis despite domain 

Computer aided pathological analysis has been the gold standard for tumor diagnosis, however domain shift is a significant problem in histopathology.  It may be caused by variability in anatomical structures, tissue preparation, and imaging processes—challenges the robustness of segmentation models. In this work, we present a framework consist of pre-trained encoder with a Swin-UNet architecture enhanced by a parallel cross-attention module to tackle the problem of adenocarcinoma segmentation across different organs and scanners, considering both morphological changes and scanner-induced domain variations. Experiment conducted on  Cross-Organ and Cross-Scanner Adenocarcinoma Segmentation challenge \footnote{https://cosas.grand-challenge.org/} dataset showed that our framework  achieved segmentation scores of 0.7469 for the cross-organ track and 0.7597 for the cross-scanner track on the final challenge test sets, and effectively navigates diverse imaging conditions and improves segmentation accuracy across varying domains. 

\end{abstract}
\section{Introduction}

Histopathology has become a transformative force, particularly in tumor diagnosis and tissue segmentation, and it can largely be attributed to the advances in deep learning. Recent development has shown tremendous promise in enhancing diagnostic precision, accelerating workflows, and enabling large-scale analysis of pathology data. However, a significant hurdle remains in the scalability and reliability of these AI-driven models: domain shift—the inherent variability in digital pathology images.  Domain shift occurs due to various factors such as variations in color, brightness, and contrast, which result from differences in staining and scanner properties. These variations result in noticeable changes in image appearance, color, resolution, and texture, which can undermine the performance of segmentation algorithms when applied to data from domains different from those they were trained on. Diversifying the training data, such as incorporating samples from multiple medical centers, can help mitigate this issue. However, there is no assurance that the trained model will fully generalize to every scenario encountered in real-world applications.

Extensive research has conducted on understanding deep neural networks trained on natural images, exploring factors such as the learned representations \cite{das2024pitvis,javed2024advancing,javed2024region,mazher2024self,stacke2019closer,de2023airogs}, the impact of loss function and batch size on these representations \cite{frosst2019analyzing,geirhos2018imagenet,qayyum2024unsupervised}, and how this knowledge can be applied to transfer learning, novelty detection, and identifying adversarial examples \cite{frosst2019analyzing,papernot2018deep,qayyum2024assessment}. Recently, Cross-Organ and Cross-Scanner Adenocarcinoma Segmentation (COSAS-2024) Challenge is organized to help overcome the above challenge address by fostering the development of resilient and generalizable semantic segmentation algorithms. The challenge specifically focuses on improving the adaptability of AI-driven diagnostic systems to domain shift, ensuring consistent performance across diverse scenarios, such as different organs and imaging devices. In this paper, we developed framework consist of pre-trained encoder with a Swin-UNet aided parallel cross-attention module for adenocarcinoma segmentation across different organs and scanners, considering both morphological changes and scanner-induced domain variations.

\begin{figure}
    \centering
    \includegraphics[width=0.65\linewidth]{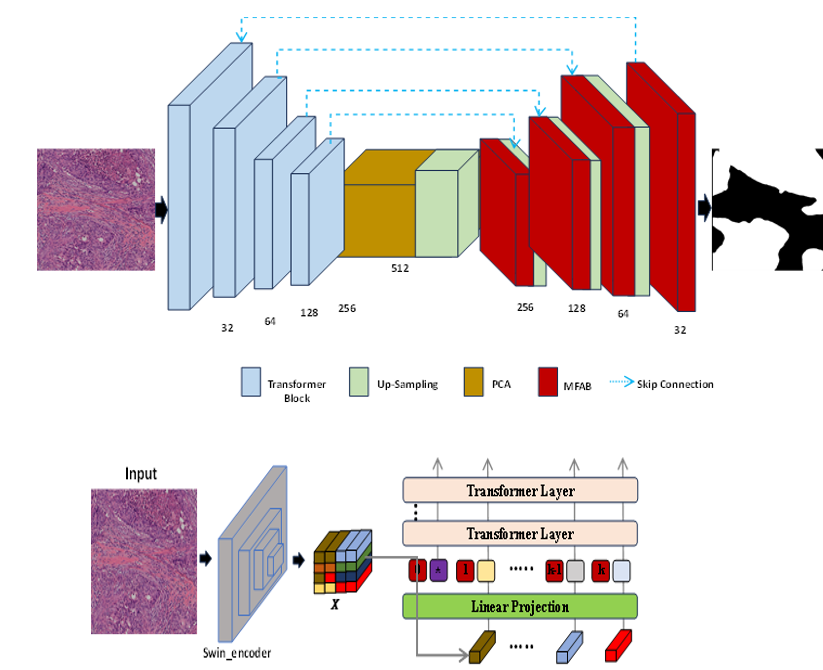}
    \caption{(top) Proposed framework for cross-organ adenocarcinoma segmentation (bottom) pretrained Transformer block used in proposed model }
    \label{fig:framework}
\end{figure}

\section{Adenocarcinoma Segmentation Framework}

Figure \ref{fig:framework} illustrates the proposed framework. For cross-domain adenocarcinoma segmentation, we adopt the Swin-UNet architecture. The primary challenge is effectively adapting generic pretrained models for segmentation tasks. We employ an encoder that mirrors the structure of recent vision model approaches, specifically Swin-UNet \cite{liu2021swin}, pretrained on the extensive ImageNet dataset. This enables the extraction of multi-scale features with long-range dependencies, reduces overfitting risk, and provides a robust initialization for the Swin-UNet, which is enhanced by a parallel cross-attention module. The Swin-UNet encoder consists of five stages. Stage 1 is a stem stage, featuring a convolutional layer for 2x down-sampling, with a 7$\times$7 kernel, padding size of 3, and stride size of 2, followed by a 2D instance normalization layer. Unlike other models, Swin-UNet uses a gradual down-sampling approach, reducing the resolution by 2x at each stage while preserving low-level details. In the second stage, a patch embedding layer with a 2×2 patch size is employed, maintaining the feature resolution at 1/4 of the original image, similar to vision transformers.

The subsequent stages follow the Swin-UNet design, with each stage consisting of a patch merging layer for 2x down-sampling and several Swin blocks for high-level feature extraction. The number of Swin blocks in stages 2 through 5 are {2, 2, 9, 2}, respectively. Feature dimensions increase quadratically across the stages, with D = {48, 96, 192, 384, 768}. The Swin blocks and patch merging layers are initialized using ImageNet-pretrained weights from the Swin-Tiny model, though the patch embedding block is not initialized with pretrained weights due to differences in patch size and input channels. The integration of a parallel cross-attention module enhances feature extraction and improves segmentation performance. A Position-wise Attention Block captures spatial dependencies between pixels globally, while a multi-scale fusion attention block (MFAB) captures channel dependencies between feature maps through multi-scale semantic feature fusion. MFAB considers both high-level and low-level feature maps, fusing their channel dependencies to obtain rich multi-scale semantic information and improve network performance."

\begin{figure}
    \centering
    \includegraphics[width=0.65\linewidth]{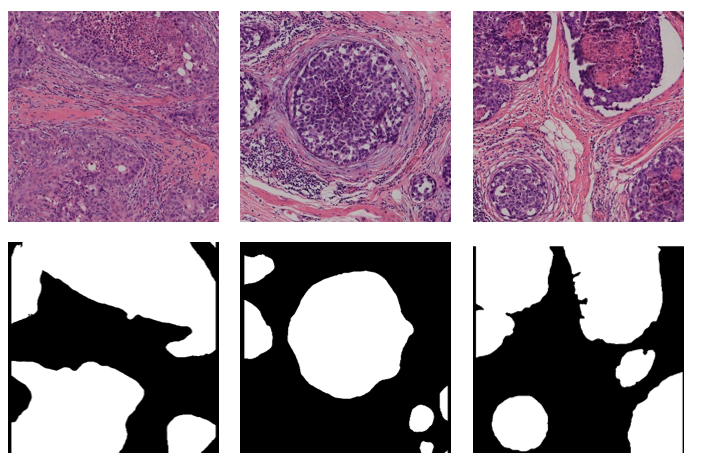}
    \caption{Segmentation map produced by our proposed model for three different samples}
    \label{fig:result}
\end{figure}
\section{Dataset}
Recently, the Cross-Organ and Cross-Scanner Adenocarcinoma Segmentation (COSAS-2024) Challenge has been organized to tackle these issues by promoting the development of robust and generalizable semantic segmentation algorithms. COSAS-2024 Challenge offers two cross-domain histology datasets, each designed to test the generalization abilities of machine learning algorithms under different conditions. Both tasks consist of 290 ROIs from human adenocarcinoma samples, with tumor lesions manually annotated. In Task 1 Cross-Organ Adenocarcinoma Segmentation dataset consists of image patches extracted from whole slide images (WSIs) representing six distinct types of adenocarcinomas, each originating from a different organ, emphasizing biological variability while using a single scanner to control for technical factors. Task 2, Cross-Scanner Adenocarcinoma Segmentation, focuses on images from invasive breast carcinoma tissues captured by six different scanners  each from a distinct manufacturer, testing models' resilience to technical variability. The variability introduced by the different scanners, such as differences in image resolution, color, contrast, and other scanner-specific factors, presents a significant challenge. The dataset is designed to test the resilience of segmentation models to domain-shifts caused by technical factors, providing a robust benchmark for assessing the adaptability of algorithms to variations in imaging devices.

\section{Experiment}
\subsection{Training and Optimization Techniques}

We trained and optimized our proposed solution on both tasks. These tasks required robust segmentation performance across varying organs and imaging scanners, which presented significant challenges for generalization. To improve model robustness, we applied a series of data augmentation techniques, including random flips, horizontal flips, and vertical flips. These augmentations helped increase the diversity of training samples and reduce overfitting, ensuring that the model learned more generalized features across different variations of the input images. For optimization, we used the Adam optimizer with a learning rate set to 0.0001. The model was trained using a batch size of 32 over 500 epochs, which allowed the network to learn from a large number of iterations, enhancing its ability to generalize to unseen data.  Given the large size of the input images, which were approximately 1500x1500 pixels, we implemented a random patch extraction strategy. During training, we extracted random 224$\times$224 patches from the original images, which made it computationally feasible to train the model while retaining significant detail in each patch. For testing and validation, we employed a sliding window approach to process the full-size images. The entire training process took approximately 5 hours. 
\subsection{Results}
To determine the final ranking for each task, the evaluation process was divided into two phases: the preliminary test phase, which accounts for 20\% of the overall ranking, and the final test phase, contributing the remaining 80\%. Our framework achieved final ranking scores of 0.7469 and 0.7597 for Task 1 and Task 2, respectively. These results demonstrate that our model performed consistently well across both tasks, delivering comparable performance in cross-organ and cross-scanner segmentation. The relatively high scores suggest that the model effectively generalizes across diverse datasets, highlighting its robustness in handling different anatomical structures (Task 1) and imaging modalities (Task 2). The slightly higher score in Task 2 indicates that the model may be better at adapting to variations introduced by different scanners, likely due to its strong feature extraction and transfer learning capabilities. This performance emphasizes the importance of robust pretraining and the use of advanced architectures like Swin-UNet with a parallel cross-attention module, enabling the model to efficiently capture multi-scale and cross-domain features. Overall, the results suggest that the proposed approach provides a solid foundation for adenocarcinoma segmentation across diverse medical imaging conditions. However, there may still be opportunities for improvement, particularly in cross-organ segmentation.

Figure \ref{fig:result} showcases the segmentation map for Cross-Organ Adenocarcinoma Segmentation generated by our proposed model. This visualization clearly demonstrates the model's ability to accurately predict adenocarcinoma regions across various anatomical structures. The figure shows that the model precisely delineates tumor boundaries, reflecting its strong capacity to localize and segment tumors effectively, even in the presence of anatomical variability. The segmentation maps further illustrate the model’s consistency across different organs, underscoring its robustness in handling diverse anatomical contexts. Additionally, the fine details of the tumors are well-preserved, showcasing the model's capability to capture subtle variations in shape and size. Overall, the visualization in Figure \ref{fig:result} validates the effectiveness of our approach in providing precise and reliable adenocarcinoma segmentation in cross-organ applications.

\section{Conclusion}
In this work, we presented an efficient framework for adenocarcinoma segmentation, addressing two challenging tasks: Cross-Organ Adenocarcinoma Segmentation and Cross-Scanner Adenocarcinoma Segmentation. By leveraging a Swin-UNet architecture with a parallel cross-attention module, proposed framework demonstrated strong performance in the COSAS-2024 challenge, achieving final ranking scores of 0.7469 for Task 1 and 0.7597 for Task 2. These results highlight the model’s ability to generalize effectively across diverse anatomical structures and imaging modalities. 

%Key to our success was the use of pretrained feature learning, effective feature extraction, and robust data augmentation strategies, which allowed the model to handle the complexities of high-resolution medical images. The integration of a sliding window approach for testing and validation enabled the processing of large images without sacrificing accuracy. While the results are promising, future work could focus on further optimizing the model for cross-organ segmentation and exploring additional techniques for improving generalization to diverse clinical environments. Overall, the proposed solution demonstrates potential for enhancing segmentation accuracy, which is crucial for real-world diagnostic applications and other medical imaging applications.
\bibliographystyle{splncs04}
\bibliography{biblo}

\end{document}